\documentclass[showpacs,showkeys,preprint]{revtex4}
\usepackage{amsmath}
\usepackage{amsfonts, amssymb}
\def\exp{\text{e}}
\def\be{\begin{equation}}
\def\ee{\end{equation}}

\begin{document}

\title{Shift in the speed of reaction diffusion equation with a cut-off: pushed and bistable fronts}
\author{R. D. Benguria and M. C. Depassier} 
  \email[Corresponding author: M. C. Depassier  ]{mcdepass@uc.cl}
\affiliation{
 Instituto  de F\'\i sica\\
	Pontificia Universidad Cat\'olica de Chile\\
	       Casilla 306, Santiago 22, Chile}

\date{\today}

\begin{abstract}
We study the change in the speed of pushed and bistable fronts of the reaction diffusion equation in the presence of a small cut-off.  We  give explicit formulas for the shift in the speed for arbitrary reaction terms $f(u)$.    The  dependence of the speed shift on the cut-off parameter is a function of the front speed and profile in the absence of the cut-off.   In order to determine 
% the power law dependence of
 the speed shift 
% on the cut-off parameter
  we solve the leading order approximation to the front profile $u(z)$ in the neighborhood of the leading edge and use a variational principle for the speed.  We  apply the general formula  to the Nagumo equation and recover the  results which have been obtained recently by geometric analysis. 
The formulas given are of general validity and we also apply them to a class of  reaction terms which have not been considered elsewhere.

\end{abstract}


\maketitle

\section{Introduction}
 
 The one dimensional reaction diffusion equation 
\be \label{eq:rd1d}
u_t = u_{xx} + f(u) \qquad \mbox{with}\qquad f(0)=f(1)=0,
\ee
has been extensively studied since the original works of Fisher \cite{Fi37}, Kolmogorov, Petrovskii and Piscounov (KPP hereafter) \cite{KoPePi37} and Zeldovich \cite{ZeFr38},  as it is the simplest equation that describes the propagation of  traveling fronts  in a variety of problems arising in physics, population dynamics,   chemistry and others.  The time evolution of localized initial conditions leads to the appearance of monotonic traveling fronts joining the  stable  $u=1$  to the  unstable $u=0$ equilibrium point. For  bistable reaction terms the traveling front may join two stable equilibria.  The convergence of initial conditions to monotonic fronts was  first  studied rigorously in \cite{KoPePi37} where it was shown  that positive  reaction terms with non vanishing derivative at the origin which in addition satisfy  $f'(u) <  f'(0)$ evolve into a monotonic front 
of speed  $c = c_{KPP} = 2
\sqrt{f'(0)}$.  
These fronts are also called pulled fronts as its speed depends on properties of the leading  edge of the front. In contrast,  fronts for which the speed depends on the full reaction term are called pushed.  These results were extended and generalized by Aronson and Weinberger \cite{ArWe78}. They  include general positive reaction functions and bistable reaction functions, which satisfy  $f(u) < 0$ for $u$ in
$(0,a)$, $f >0$ on $(a,1)$ with $\int_0^1 f(u)\, du >0$.  It was proved in 
\cite{ArWe78} that  sufficiently localized  initial conditions  evolve into a monotonic travelling front $u =
U(x-c\,t)$ joining the stable state $u=1$ to the state  $u=0$.  For reaction terms which are positive in $(0,1)$ 
 there is continuum of values of $c$ for which a monotonic
front exists and the system evolves into the front of minimal
speed. The minimal speed, $c$  satisfies $2 \sqrt{f'(0)} \leq c \leq 2 \sup \sqrt{f(u)/u}$ thus generalizing the condition given in \cite{KoPePi37}.  For bistable reaction terms  there is a single isolated value
of the speed for which the monotonic front exists.

The reaction diffusion equation (\ref{eq:rd1d}) is the simplest model exhibiting front propagation.  In many problems other effects such as  density dependent diffusion, memory effects, convective terms and other  are important and have been studied as well.   A different effect was studied by Brunet and Derrida \cite{BrDe97}  who wished to model the effect of additive noise and the finiteness in the number of diffusing particle on the diffusing front.  They conjectured that these effects can be modeled by the classical reaction diffusion equation (\ref{eq:rd1d}) introducing a  small cut-off at the edge of the propagating front. This conjecture was validated numerically in \cite{BrDe97}   for a KPP type reaction function.  Using perturbation methods they calculated  the leading order correction to the speed,
\be \label{eq:Brunet}
c \approx 2 -\frac{\pi^2}{(\log \epsilon)^2},
\ee
result obtained solving the equation for traveling fronts $U(z=x- c t)$ of (\ref{eq:rd1d}) asymptotically. Recently these results   were established rigorously  in  \cite{MuMyQu11} where a reaction diffusion equation of KPP type with noise is studied and  where it is proved that the speed is effectively (\ref{eq:Brunet}) to leading order. 
The study of the speed for a KPP type reaction term with a cut-off was studied rigorously in  \cite{DuPoKa07} using geometric singular perturbation theory and a blow-up method and in \cite{BeDe07, BeDeLo12} through a variational approach.
While the speed shift due to the cut-off equation (\ref{eq:Brunet}) is valid for all KPP fronts with a cut-off, for pushed and bistable fronts the speed, and the change in the speed due to the cut-off  depends on the full reaction term.  The shift in the speed was first found to depend on powers of the cut-off \cite{KeNeSa98} for the Nagumo equation with the exact value depending on the reaction term itself.  Small perturbations to the reaction term were  shown to  have a marked effect on the speed \cite{PavanSa02}.   
A variational approach was used in \cite{MeCaZe05} and in \cite{BeDeHa07} where the  correct power law behavior is obtained but no the magnitude of the shift.   
 The method of geometric singular perturbation has been  applied to calculate rigorously  the  exact leading order behavior of  shift in the speed due to a cut-off  for the  Nagumo equation $f(u) = u (1-u)(u-a)$ \cite{DuPoKa10} in the bistable regime $0<a<1/2,$ and to the reaction term $f(u) = u^m(1-u)$ \cite{Po11}.

The purpose of this work is to give a general expression for the shift in the speed for bistable and pushed fronts for arbitrary reaction terms.  We use an integral variational principle for the speed and find that two cases have to be distinguished,  depending on the vanishing or non vanishing derivative of the uncut reaction term at the origin. We give a general formula for each case, both of which depend on the solution to the original front. The shift in the speed due to the cut-off  is found to depend on the cut-off parameter and on the speed and rate of approach to the leading edge of the front  in the absence of the cut-off.

We show that the shift in the speed is given by
\be\label{eq:caso1}
\Delta c =   -  K  f'(0)    \frac{ \epsilon^{2 + c_0/k_2}}{(2 + c_0/k_2)^{1+c_0/k_2} } \text{     when   } f'(0)\neq 0.
\ee
where  $c_0$ is the speed of the front in the absence of a cut-off, $k_2$ is the rate of approach of the front to the leading edge in the absence of a cut-off, that is $u\approx \exp^{ k_2 z}$ as $z\rightarrow \infty$ and is given by $ k_2 = -c_0/2 - \sqrt{c_0^2 - 4 f'(0)}/2. $  $K$ is a constant which depends on an integral of the exact solution of the front without cut-off.   If the derivative of the reaction term vanishes at the origin we show that the shift in the speed is obtained integrating
\be \label{eq:caso2}
\frac{d \Delta c}{d\epsilon} = - K  \left(  \frac{f(\epsilon)}{\epsilon} \right).
\ee

In  Section 2  we describe the problem and state known results which are needed to obtain the shift in the speed. In this section the main results are derived. In Section 3  we apply the general formula to cases where the exact solution in the absence of a cut-off is known, allowing then for the complete determination of the shift in the speed.  We conclude indicating possible extensions of these results.

\section{Speed of the fronts}

In this section we begin recalling properties of the traveling fronts  of the reaction diffusion equation. It was shown in \cite{KoPePi37,ArWe78} that  for a a wide class of reaction terms $f(u)$   the solution  of (\ref{eq:rd1d}) starting from  a sufficiently a localized initial condition evolves into monotonic traveling front $u(x,t) = U(z=x - c t)$ which  obeys the ordinary differential equation
\be \label{eq:ODE}
U_{zz} + c U_z + f(U) = 0,  \quad \lim U_{z \rightarrow
-\infty} = 1 ,  \quad\text{and}\quad \lim U_{z \rightarrow \infty} = 0.
\ee

 In this work we consider reaction terms of bistable type, that is 
 $f(u) < 0$ for $u$ in
$(0,a)$, $f >0$ on $(a,1)$ with $\int_0^1 f(u)\, du >0$, and  pushed fronts, that is fronts for which $f > 0$ in $(0,1)$ but whose  speed is greater than the KPP value, $ 2 \sqrt{f'(0)}.$
   In both  cases the front approaches $U=0$ exponentially with the decay rate \cite{ArWe78}
 $$
U  \approx   \exp^{ -  [ c + \sqrt{c^2 - 4 f'(0)}\, ] z /2 } \qquad \text{as}\quad {z\rightarrow \infty}
 $$
 In previous work \cite{BeDe96} we showed that the speed of the front satisfies the integral variational principle
\be 
\label{eq:var}
c^2\,=  \sup \left( 2\, {{\int_0^1  f(U)\, g(U)\, du}\over{\int_0^1
 (-g^2(U)/g'(U)) dU}} \right),
\end{equation}
where the supremum is taken over all positive decreasing
functions $g$
in $(0,1)$ for which the integrals exist.
Moreover there is always a maximizing $g$  for bistable reaction terms  and for oushed fronts.
 The optimal $g$, say $\hat g$,  is the solution of
 \be \label{eq:ghat}
\frac{ \hat g'}{\hat g} = -\frac{c}{p} \quad\text{where}\quad p(U) =-\frac{d U}{d z}
\ee
 This variational principle is the starting point for our derivation. Since we are interested in bistable and pushed fronts, we know that a maximizing $g$ exists \cite{BeDe96} and the 
 Feynman-Hellman theorem holds. The Feynman-Hellman theorem states that if  the reaction term $f$ depends on a parameter $\alpha$ (i.e., $f=f(U,\alpha)$), then
 \begin{equation}
\frac{\partial c^2}{\partial \alpha} = 2 \frac{\int_0^1  \frac{\partial f}{\partial \alpha}(U,\alpha)\, \hat g(U,\alpha)\, dU}{\int_0^1
 (-\hat g^2/\hat g_U) dU},
\label{eq:fh}
\end{equation}
where $\hat g(U,\alpha)$ is the function (unique up to a multiplicative constant) that yields the maximum in (\ref{eq:var}) at the given parameter $\alpha$. We use a subscript to denote the partial derivative with respect to the corresponding argument. Notice that the Feynman-Hellman theorem holds only if the maximum is attained, which is not the case for KPP fronts.

We apply now the Feynman-Hellman theorem to  pushed or bistable reaction term $f(u)$ to which we appply a cut-off, that is the reaction term becomes   $ f(u) \Theta(u-\epsilon)$ (here, $\Theta(x)$ denotes the Heaviside step function). The Feynman-Hellman theorem tells us that
\begin{equation} \label{eq:FeynHell}
\frac{\partial c^2}{\partial \epsilon} = 2 \frac{\int_0^1  \frac{\partial f(U) \Theta(U-\epsilon)}{\partial \epsilon}\, \hat g(U,\epsilon)\, dU}{\int_0^1
 (-\hat g^2/\hat g') dU},
= - 2 \frac{f(\epsilon)\hat g(U=\epsilon,\epsilon)}{\int_0^1
 (-\hat g^2/\hat g') dU}
\end{equation}
In the expression above $\hat g = \hat g(U,\epsilon)$ is the optimizing function for the speed of the front with the reaction term $f(U)\Theta(u-\epsilon)$. Let us call $c_0$ and $U_0(z)$ the speed and the front profile in the absence of the cut-off,  which  from here on we  call the unperturbed problem.
Since the cut-off is small we expect the shift in the speed to be small as well. We write then
$$
c = c_0 + \Delta c
$$
so that,  to leading order,   (\ref{eq:FeynHell}) implies
\be \label{eq:main}
\frac{d \Delta c}{d\epsilon} =  - K  f(\epsilon)\hat g(\epsilon,\epsilon) \quad \text{with} \quad  K = \frac{1}{c_0 \int_0^1 (-\hat g^2/\hat g_U) dU}.
\ee

 In order to calculate $\hat g(\epsilon,\epsilon)$ we notice that equation (\ref{eq:ghat}) can be solved explicitly in terms of the space variable $z$. Replacing $p$ by its definition we observe that $\hat g$  is given by 
 \be \label{eq:gz}
\hat g = \exp^{c z}
\ee
up to a multiplicative constant constant, which we can set  equal to 1 due to the
 translation invariance of equation (\ref{eq:ODE}) and invariance to scaling in $g$ of the variational principle (\ref{eq:var}).  It is  convenient to calculate $\hat g(u=\epsilon,\epsilon) $ by solving the problem in the space variables,  for which  the leading order profile  $U(z)$ including the cut-off must be obtained.

In the region   $0<U \le \epsilon$ where the reaction term is zero, which we shall call the inner region ( the leading edge of the front) the front profile  satisfies
$$
U_{1zz} + c U_{1z} = 0, \text{    with  }    \lim U_1  =0  \quad\text{as}\quad {z\rightarrow\infty},
$$
the solution of which is
$
U_1(z) = A \exp^{- c z}.
$
If we let $c= c_0 + \Delta c$, to leading order we may write
\be
\label{eq:U1}
U_1(z) = A \exp^{- c _0 z}.
\ee
Let  us call $z=z^*$  the spatial coordinate where $U_1=\epsilon.$   It  follows  from  (\ref{eq:gz}) that
\be
\label{eq:gzstar}
\hat g (U=\epsilon,\epsilon) = \exp^{c_0 z^* } = \frac{A}{\epsilon},
\ee
where the constant $A$ has to be determined. 

Far from the cut-off, which we call the outer region, $\epsilon <  U \leq 1$  the solution is in leading order   the unperturbed solution $U_0(z)$.  
It was proved in \cite{ArWe78}  that the front approaches the leading edge as
\be
\label{eq:Uasymp}
 U_0( z ) =  \exp^{ k_2 z } \quad \text{as} \quad {z\rightarrow\infty},  \quad\text{with   }   k_2 =  - \frac{c_0}{2} -  \frac{1}{2} \sqrt{c_0^2 - 4 f'(0)}.
\ee

Moreover, since the  profile for the front  is a solution of the second order differential equation $U_{zz} + c U_z + f(U)=0$  we know that the profile and its derivative are continuous.  Therefore the solution and its derivative in each region has to be matched.  We find that  two cases have to be distinguished according to the value of $f'(0).$ 

\subsection{The case $f'(0)$}

In this case the rate of exponential approach to the leading edge of the unperturbed  profile is $k_2 = -c_0 $, so that 
$
 U_0( z ) \approx   \exp^{ -c_0 z } 
$ as  ${z\rightarrow\infty}$
which  matches smoothly  to the leading order inner solution $U_1$ choosing  $A=1.$
In this case then,
$$
\hat g (U=\epsilon,\epsilon)  = \frac{1}{\epsilon}, \quad\text{when}\quad  f'(0)=0.
$$

It is worth mentioning that due to translation  invariance, we could have set $
 U_0( z ) \approx  \text{constant}  \times  \exp^{ -c_0 z }
$ as  ${z\rightarrow\infty}$ which would lead to a different value for $A$. But since the problem is invariant under translations  in $z$ and the  variational principle (\ref{eq:var})  and equation (\ref{eq:main}) are invariant with respect to a scaling in $g$ such a constant  cancels out in the final result. Therefore, with no loss of generality, we may set it equal to 1.

Using this result in (\ref{eq:main}) we obtain
\be\label{eq:res1}
\frac{d \Delta c}{d\epsilon} =  - K  \frac{f(\epsilon)}{\epsilon}.
\ee
Finally,  to leading order, the main contribution to $K$ arises from the unperturbed solution $g_0(U)$. The  correction due to the cut-off will only affect higher order terms  to the speed shift, that is, we may approximate
\be\label{eq:res1a}
 K = \frac{1}{c_0 \int_0^1 (-\hat g_0^2/\hat g_0') dU}  + \text{small corrections.} 
 \ee
The shift to the speed in this case is obtained integrating (\ref{eq:res1}).

\subsection{The case $f'(0) \neq 0$}

In this case the inner solution (\ref{eq:U1}) cannot be matched directly to the outer solution as the value of  $k_2$  does not allow smooth matching of the profiles. This indicates the existence of a transition layer in the region  $ U \gtrsim \epsilon$

In this region, for sufficiently low values of $\epsilon$  the reaction term can be approximated by the linear form  $f(U) \approx U f'(0).$  In leading order the front satisfies the equation
$$
U_{2zz} + c_0 U_{2z} + U f'(0) = 0,
$$
the solution of which is
$$
U_2(z) = B_1 \exp^{k_1 z} + B_2 \exp^{k_2 z}
$$
where 
$$
k_1 =  - \frac{c_0}{2} + \frac{1}{2} \sqrt{c_0^2 - 4 f'(0)}, \qquad k_2 =  - \frac{c_0}{2} -  \frac{1}{2} \sqrt{c_0^2 - 4 f'(0)}, 
$$
We must match $U_1$ to $U_2$ at $U=\epsilon$ to obtain the coefficients $B_1$ and $B_2$ in terms of $A$.  Let us call $z^*$ the value of $z$ when $U=\epsilon$ as before. The matching conditions are then
$$
U_2 = \epsilon \text{    and   } U_{1z} = U_{2z}   \text{     at   } z^* = \ln \left( \frac{A}{\epsilon} \right)^{1/c_0}.
$$
The solution of this system yields
$$
B_1 = \epsilon \frac{k_2 + c_0}{k_2-k_1} \exp^{- k_1 z^*},   \qquad B_2 =   - \epsilon \frac{k_1 + c_0}{k_2-k_1} \exp^{- k_2 z^*}
$$
Replacing the expressions for $k_1, k_2$ and $z^*$ this is
$$
B_1 =  \frac{\epsilon [  \sqrt{c_0^2 - 4 f'(0)}- c_0 ]}{2  \sqrt{c_0^2 - 4 f'(0)}} \,  \left( \frac{\epsilon}{A}\right)^{k_1/c_0}  \quad  B_2 =  \frac{\epsilon [ c_0 + \sqrt{c_0^2 - 4 f'(0)}\, ] }{2  \sqrt{c_0^2 - 4 f'(0)}} \,  \left( \frac{\epsilon}{A}\right)^{k_2/c_0}.
$$
Having matched this intermediate solution to the inner solution  $U_1$ we now match $U_2$ to the outer solution valid farther from the cut-off.   This matching condition follows from   
(\ref{eq:Uasymp}),
 \be
U_2(z)  \rightarrow  \exp^{k_2  z} \quad\text{as}\quad z\rightarrow - \infty.
\ee
This implies $B_2 = 1$ and we obtain
$$
A =  \epsilon^{1 + c_0/k_2} \left[   \frac{c_0 + \sqrt{c_0^2 - 4 f'(0)}\,  }{2  \sqrt{c_0^2 - 4 f'(0)}}\right]^{c_0/k_2}.
$$
Replacing this value of $A$ in (\ref{eq:gzstar}) we obtain
$$
\hat g (u=\epsilon,\epsilon)  = \epsilon^{c_0/k_2} \left[   \frac{c_0 + \sqrt{c_0^2 - 4 f'(0)}\,  }{2  \sqrt{c_0^2 - 4 f'(0)}}\right]^{c_0/k_2},  \quad\text{when}\quad  f'(0)\neq 0.
$$
To calculate the shift in the speed,   we go back to (\ref{eq:main}), with $f(\epsilon) = \epsilon f'(0)$, and using the leading order value (\ref{eq:res1a}) for $K$.  Integrating with respect to $\epsilon$ we obtain the shift in the speed
\be\label{eq:res2}
\Delta c =   -  K  f'(0)   \left[   \frac{c_0 + \sqrt{c_0^2 - 4 f'(0)}\,  }{2  \sqrt{c_0^2 - 4 f'(0)}}\right]^{c_0/k_2} \, \frac{ \epsilon^{2 + c_0/k_2}}{2 + c_0/k_2} .
\ee 
with $K$ given in (\ref{eq:res1a}). Replacing the value of $k_2$ in the expression above, the shift in the speed  can be written in the  compact form  (\ref{eq:caso1}).

Equations (\ref{eq:res1}) and (\ref{eq:res2}) (or its equivalent (\ref{eq:caso1})) together with (\ref{eq:res1a}) constitute our main result.

\section{Examples}

We will apply the results obtained in the previous section to the Nagumo equation, 
\begin{equation} f(u) = u (1-u) (u - a)
\end{equation}
for which an exact solution of the unperturbed case is known and $K$ can be calculated explicitly.  The shift in the speed due to the cut-off the speed has been calculated  rigorously \cite{DuPoKa10} with a geometric approach.

For $0<a<1/2$ this is a bistable reaction term. For  $-1/2<a<0$ it gives rise to a pushed front. The case $a=0$ is an example with vanishing derivative at the origin. 
The speed without the cutoff is given by
\begin{equation} c_0 = \frac{1}{ \sqrt{2}} - a \sqrt{2} \end{equation}
which is obtained from the variational principle (\ref{eq:var}) with the trial function \cite{BeDe96}
\begin{equation}
\hat g_0(U) = \left( \frac{{1-U}}{ U} \right)^{1 -
2 a}.\end{equation}
For this reaction term $f'(0) = - a$. The value of $K$ is
\begin{equation}
K = \left[c_0 \int_0^1 (-g_0^2/g_0') dU \right]^{-1} =
 \frac{\sqrt{2}\,\Gamma(4)}{\Gamma(1+ 2 a)\Gamma(3 - 2 a)}.
\end{equation}

In the bistable and in the pushed regime replacing the value of $c_0$ and $f'(0)= -a$,  in (\ref{eq:res2})  we obtain
$$
\Delta c = \frac{\sqrt{2}\,\Gamma(4) a }{\Gamma(1+ 2 a)\Gamma(3 - 2 a)} \,\frac{\epsilon^{1 + 2 a}}{(1+ 2 a)^{2 a}}.
$$
in agreement with the result of \cite{DuPoKa10}.

When $a=0$, $f'(0)=0$ and the shift is obtained using (\ref{eq:res1}). In this case  $f(\epsilon) = \epsilon^2$ to leading order,  $c_0 = 1/\sqrt{2}$ and $g_0(u) = (1-u)/u$ so that 
$$
K =   3 \sqrt{2}
$$
The shift in the speed is the solution of
$$
\frac{d \Delta c}{d\epsilon} =  - 3\sqrt{2} \epsilon
$$
that is,
$$
 \Delta c = -  \frac{3}{\sqrt{2}} \epsilon^2
 $$
 in agreement with the result obtained in (\cite{Po11}) by geometric analysis.
 
 As a third  example take the family of reaction terms also considered in \cite{Po11}
 $$f_m(u) = u^m (1-u).
 $$
 The effect of the cut-off is to shift the speed according to (\ref{eq:res1}) so that 
 $$
 \frac{d \Delta c}{d\epsilon} =  - K  \frac{\epsilon^m}{\epsilon}.
 $$
 from where we recover theorem 1.1 of \cite{Po11} obtained via geometric analysis,
 $$
 \Delta c = - K \frac{ \epsilon^m}{m}
 $$
 with the identification $\gamma_m = -K/m$. We cannot compute explicitly the value of $K$ since the unperturbed solution is not known except in the case $m=2$ already described above.

 Finally we apply the results to the class of exactly solvable reaction terms, which have not been considered elsewhere, 
 $$
 f(u) = f'(0) \left( u + \frac{n+1-\lambda}{\lambda-1}  u^n -\frac{n}{\lambda-1} u^{2 n -1} \right),
 $$
with $ \lambda\neq 1\,  \lambda <2,\, n>1,$ and $ \lambda+ n >1.$

 The traveling front solution to (\ref{eq:ODE}) is given by \cite{BeDe95}
 \be\label{eq:U0n}
 U_{0}(z) = \frac{ \exp^{- a_1 z}}{ ( 1 +  \exp^{- (n-1) a_1 z})^{1/(n-1)}}, \,\text{where}\quad  a_1 = \sqrt{\frac{f'(0)}{\lambda-1}}
 \ee
  which travels with speed
$$  c_0 = \lambda      \sqrt{\frac{f'(0)}{\lambda-1}}.
 $$
 The phase space solution for this front is 
 $$
 p_{0 } (U) = - \frac{dU}{d z}(U) = a_1 U (1 - U^{n-1}).
 $$
 For the values of $\lambda$ specified, this solution corresponds to pushed or bistable fronts \cite{BeDe95}. 
Using equations (\ref{eq:gz}) and (\ref{eq:U0n})  we obtain the optimal trial function for the exact front $\hat g_0(U),$
$$
g_0(U) = \exp^{c_0 z} = \left( \frac{1- U^{n-1}}{U^{n-1}}\right)^{\lambda /(n-1)}.
$$
In order to calculate the speed shift we need to calculate
\be\label{eq:Kejemplo}
 \frac{1}{K} =c_0 \int_0^1 - \frac{\hat g_0^2}{ \hat g_0'} dU = \int_0^1 g_0(U) p_0(U) d\,U = 
a_1  \frac{ \Gamma (2 + \frac{\lambda}{n-1}) \Gamma ( \frac{2-\lambda}{n-1})}{ (n+1)  \Gamma( \frac{n+1}{n-1}) },
 \ee
 where we  used (\ref{eq:ghat}) as an intermediate step. Replacing this result in equations (\ref{eq:caso1}) or its equivalent (\ref{eq:res2}), and using $k_2 = - a_1$ we obtain the final expression for the shift in the speed to be
 $$
 \Delta c = - K \frac{\epsilon^{2-\lambda}}{(2-\lambda)^{1-\lambda}}
 $$
 with $K$ given in (\ref{eq:Kejemplo}). The breakdown of this expression at $\lambda=2$ obeys to the transition from a pushed to a KPP front that occurs at $\lambda =2$  \cite{BeDe95} for which a different approach must be used.  The Nagumo equation  is a special case  with $n=2$, $\lambda = 1 - 2 a.$ 
 
 \section{Summary}
 
 We studied the change in speed of a reaction diffusion front due to a cut-off in the case when the original front without a cut-off is bistable or pushed. We used a variational principle in order to calculate the shift in the speed. We distinguished two cases, according to the behavior at the leading edge of the reaction term, that in which the derivative of the reaction term vanishes at the origin  and those in which it does not. This last case includes bistable and positive non-KPP fronts. we have only considered one type of cut-off, a complete cut-off at the edge of the front. While we considered only the simplest classical reaction diffusion equation (\ref{eq:rd1d}) the method we have used can be extended to generalized reaction diffusion equations for which a variational principle for the speed  has been formulated \cite{BeDe02,BeDeMe04}.   While  a Heaviside cut-off has been the most widely studied case, a gradual cut-off is also possible and has been studied in  \cite{Po12}. It is an open problem to study whether such type of cut-off can be studied via the variational approach.

 \section*{Acknowledgements}
 This work has been partially supported by Iniciativa Cientifica Milenio, ICM (Chile), through the Millenium Nucleus RC120002 ``F'sica Matem‡tica'' and  Fondecyt (Chile) projects    1100679 and  1120836.

%%%%%%%%%%%%%%%%%%%%%%%%%%%%%%%%%%%%%%%%%%%%%%%%%%%%%%%%%%%%
 
\end{document}